\documentclass{egpubl}
\usepackage{egsr2022}

\usepackage[T1]{fontenc}
\usepackage{dfadobe}  

\biberVersion
\BibtexOrBiblatex
\usepackage[backend=biber,bibstyle=EG,citestyle=alphabetic,mincitenames=1,maxcitenames=2,maxbibnames=100,giveninits=true,backref=true,sorting=anyt,doi=true]{biblatex}
\AtEveryBibitem{\clearfield{pages}}
\electronicVersion
\PrintedOrElectronic

\ifpdf \usepackage[pdftex]{graphicx} \pdfcompresslevel=9
\else \usepackage[dvips]{graphicx} \fi

\usepackage{egweblnk} 
\usepackage{amsmath}

\usepackage[nameinlink,capitalise]{cleveref}
\usepackage{relsize}
\usepackage{tikz}
\usepackage[normalem]{ulem}
\usepackage[outline]{contour}
\contourlength{0.1em}%
\usepackage{booktabs}
\usepackage{array}
\usepackage{makecell}
\usepackage{multirow}

\newcommand{\Dist}{t}
\newcommand{\Dir}{\boldsymbol{\omega}}
\newcommand{\DirX}{\Dir}
\newcommand{\DirY}{\Dir'}
\newcommand{\Trans}{\mathrm{Tr}}
\newcommand{\FFPdf}[1]{\mathrm{p}\!\left(#1\right)}
\newcommand{\Phase}{\rho}
\newcommand{\ExtCoeff}{\sigma}

\newcommand{\PathSpace}{\mathcal{P}}
\newcommand{\DirSpace}{\Omega}
\newcommand{\DistSpace}{\mathcal{T}}

\newcommand{\ExitP}{\mathrm{P}_\mathrm{exit}}
\newcommand{\ExitUp}{\ExitP^\uparrow}
\newcommand{\ExitDown}{\ExitP^\downarrow}

\newcommand{\DeltaZ}{{\Delta z}}

\newcommand{\DistAt}[1]{\Dist_{#1}}
\newcommand{\DirAt}[1]{\Dir_{#1}}
\newcommand{\DirI}{\DirAt{i}}
\newcommand{\DirO}{\DirAt{o}}
\newcommand{\DeltaZAt}[1]{{\DeltaZ_{#1}}}
\newcommand{\ExtCoeffAt}[1]{{\ExtCoeff_{#1}}}

\newcommand{\Abs}[1]{\lvert #1 \rvert}

\newcommand{\ZComponent}[1]{\left(#1\right)_z}

\newcommand{\SlabD}{L}

\newcommand{\HeightPdf}[1]{h_{#1}}

\newcommand{\MultiIndex}[3]{#1_{#2,#3}}
\newcommand{\seqA}[2]{\MultiIndex{a}{#1}{#2}}
\newcommand{\seqB}[2]{\MultiIndex{b}{#1}{#2}}

\definecolor{colorA}{HTML}{e30021}
\definecolor{colorB}{HTML}{edcc00}

\crefname{algocf}{Alg.}{Algs.}

\usepackage[linesnumbered,scleft,lined,boxed,commentsnumbered,ruled,noend]{algorithm2e} 
\SetAlFnt{\small}
\SetAlCapFnt{\small}
\SetAlCapNameFnt{\small}
\SetAlCapHSkip{0pt}
\IncMargin{-\parindent}
\SetKwProg{Fn}{function}{}{}
\SetKwProg{Class}{class}{}{}

\SetCommentSty{mycommfont}
\newcommand{\ClassName}[1]{\mathrm{#1}}
\SetKw{Def}{def}
\SetKw{Break}{break}

\definecolor{OliveGreen}{HTML}{3C8031}

\newcommand{\PaperTitle}{A Position-Free Path Integral for Homogeneous Slabs and Multiple Scattering on Smith Microfacets}

\newlength{\bigw}
\newlength{\insetw}
\newlength{\gridw}

\newcommand{\citet}[1]{\textcite{#1}}

\title[\PaperTitle]{\PaperTitle}

\author{paper1053}

\hypersetup{draft}

\author[B. Bitterli \& E. d'Eon]
{\parbox{\textwidth}{\centering Benedikt Bitterli\orcid{0000-0002-8799-7119}
        and Eugene d'Eon\orcid{0000-0002-3761-2989} 
        }
        \\
{\parbox{\textwidth}{\centering NVIDIA}
}
}

\begin{document}

\pagestyle{plain}

\teaser{
 \centering
    \newcommand{\insetratio}{0.3185840707964602}%
    \setlength{\bigw}{0.37540268456375836\textwidth}%
    \setlength{\insetw}{\insetratio\bigw}%
    \setlength{\gridw}{0.37540268456375836\textwidth}%
    \setlength{\gridw}{0.37540268456375836\textwidth}%
    \begin{tikzpicture}[x=\gridw,y=0.6371681415929203\gridw,every text node part/.style={align=center}]
        \node[anchor=north west] at (0, 0) {\includegraphics[width=\bigw]{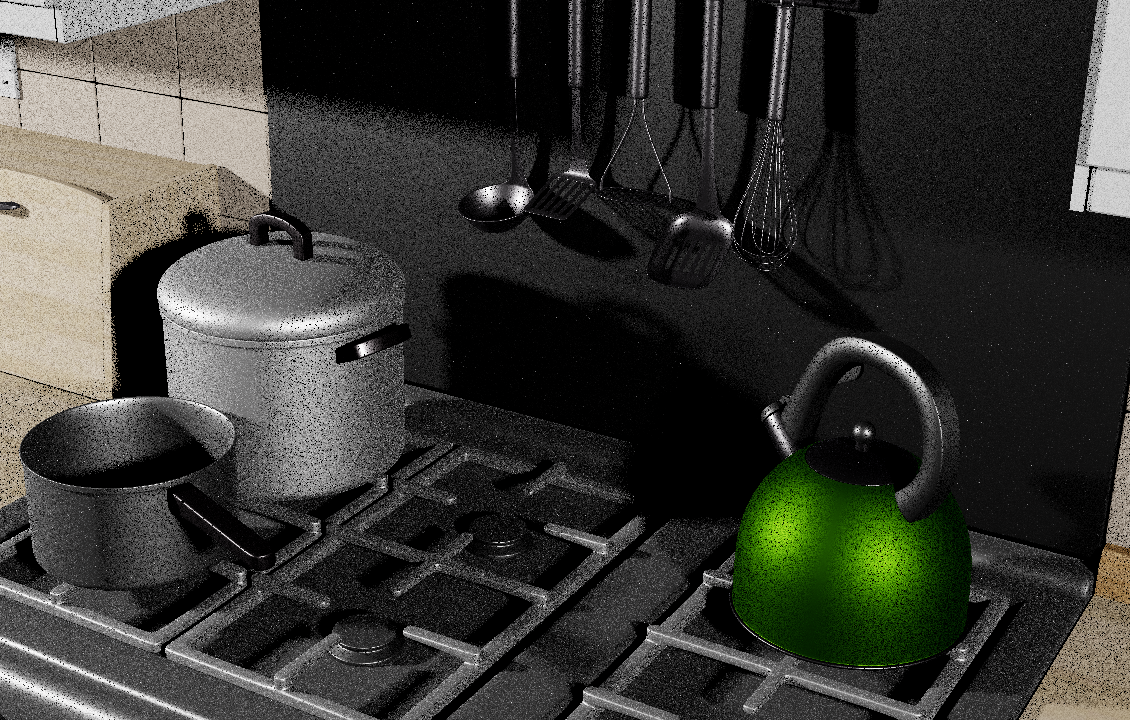}};
        \node[anchor=north west] at (1 + 2*\insetratio, 0) {\includegraphics[width=\bigw]{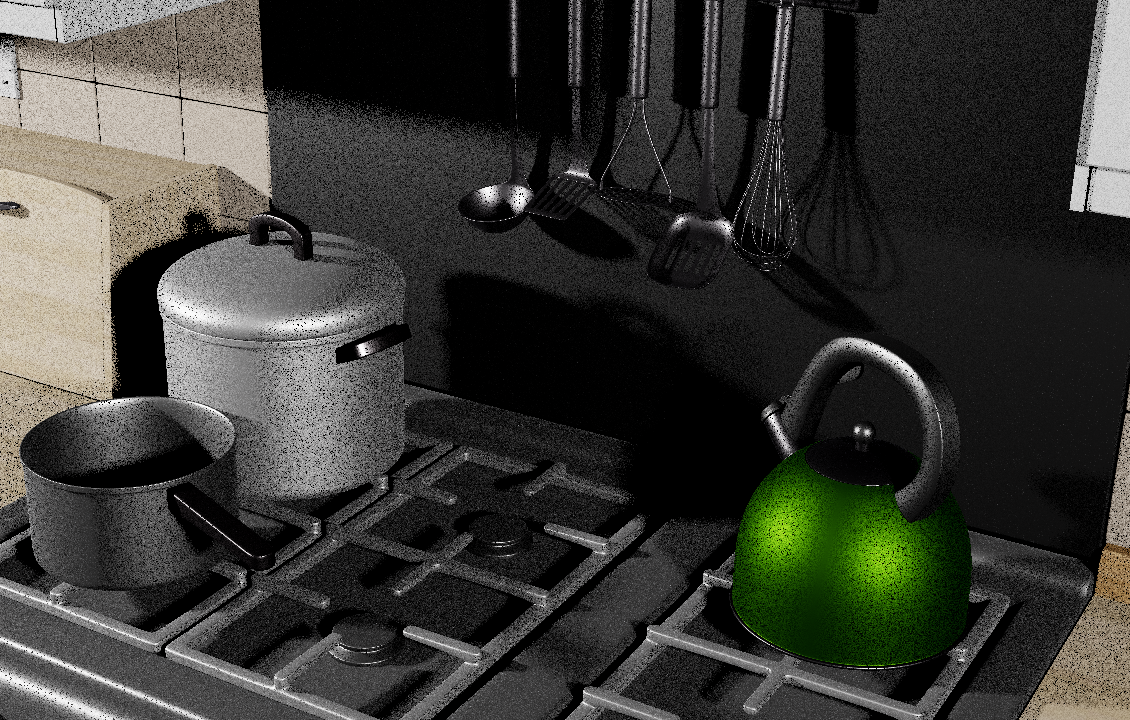}};
        \draw[draw=colorA,line width=1.5] (0.19557522123893806, -0.32083333333333336) rectangle ++(0.08849557522123894, -0.1388888888888889);
        \draw[draw=colorA,line width=1.5] (1 + 2*\insetratio + 0.19557522123893806, -0.32083333333333336) rectangle ++(0.08849557522123894, -0.1388888888888889);
        \draw[draw=colorB,line width=1.5] (0.4946902654867257, -0.7791666666666667) rectangle ++(0.08849557522123894, -0.1388888888888889);
        \draw[draw=colorB,line width=1.5] (1 + 2*\insetratio + 0.4946902654867257, -0.7791666666666667) rectangle ++(0.08849557522123894, -0.1388888888888889);
        \node[anchor=north west] at (1 + 0*\insetratio, -0.0) {\includegraphics[width=\insetw]{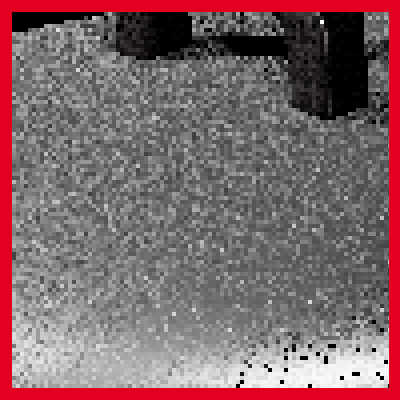}};
        \node[anchor=north west] at (1 + 0*\insetratio, -0.5) {\includegraphics[width=\insetw]{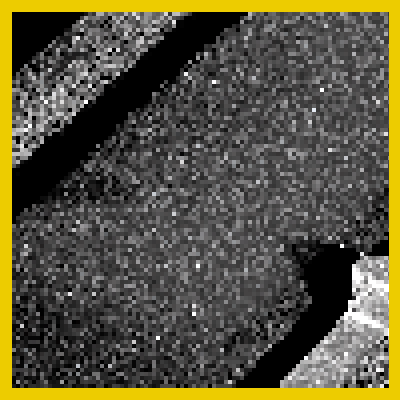}};
        \node[anchor=north west] at (1 + 1*\insetratio, -0.0) {\includegraphics[width=\insetw]{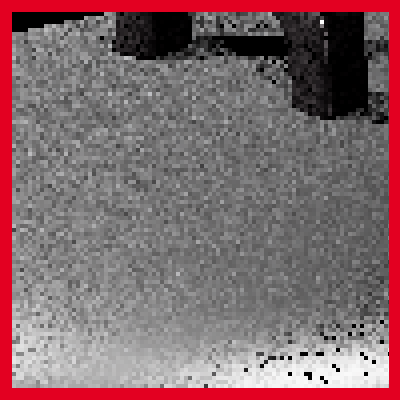}};
        \node[anchor=north west] at (1 + 1*\insetratio, -0.5) {\includegraphics[width=\insetw]{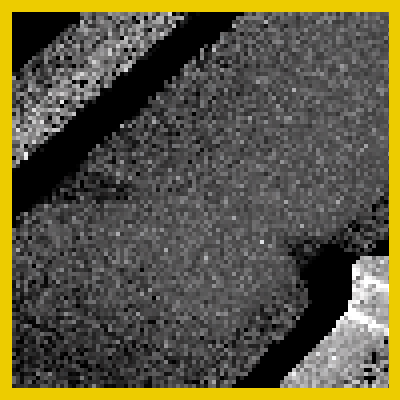}};
        \node[anchor=north] at (1 + 0.02 + 0.5*\insetratio, -0.4 - 0.125) {\footnotesize\bfseries\contour{black}{\color{white}\citeauthor{Dupuy:2016:Additional}}};
        \node[anchor=north] at (1 + 0.02 + 1.5*\insetratio, -0.4 - 0.125) {\footnotesize\bfseries\contour{black}{\color{white}Ours}};
        \node[anchor=north] at (1 + 0.02 + 0.5*\insetratio, -0.4 - 0.195) {\footnotesize\bfseries\contour{black}{\color{white}MSE: 0.0641}};
        \node[anchor=north] at (1 + 0.02 + 1.5*\insetratio, -0.4 - 0.195) {\footnotesize\bfseries\contour{black}{\color{white}MSE: 0.0437}};
        \node[anchor=north west] at (0.02, -0.02) {\bfseries\contour{black}{\color{white}\citeauthor{Dupuy:2016:Additional}}};
        \node[anchor=north west] at (0.02, -0.02 - 1.5em) {\small\bfseries\contour{black}{\color{white}108ms}};
        \node[anchor=north west] at (1 + 2*\insetratio + 0.02, -0.02) {\bfseries\contour{black}{\color{white}Ours}};
        \node[anchor=north west] at (1 + 2*\insetratio + 0.02, -0.02 - 1.5em) {\small\bfseries\contour{black}{\color{white}110ms}};
    \end{tikzpicture}\\[-2mm]%
  \caption{We show a scene containing a number of different microfacet conductors under direct lighting. All microfacets feature multiple scattering and use the GGX distribution with varying roughnesses ($\alpha\in [0.05,0.5]$). Compared to the baseline method of Dupuy et al.~\cite{Dupuy:2016:Additional} (left), our position-free bidirectional estimator of multiple scattering (right) leads to reduced variance at equal cost.}
  \label{fig:teaser}
}

\maketitle

\begin{abstract}
   We consider the problem of multiple scattering on Smith microfacets. This problem is equivalent to computing volumetric light transport in a homogeneous slab. Although the symmetry of the slab allows for significant simplification, fully analytic solutions are scarce and not general enough for most applications. Standard Monte Carlo simulation, although general, is expensive and leads to variance that must be dealt with.

    We present the first unbiased, truly position-free path integral for evaluating the BSDF of a homogeneous slab.  We collapse the spatially-1D path integral of previous works to a position-free form using an analytical preintegration of collision distances.  Evaluation of the resulting path integral, which now contains only directions, reduces to simple recursive manipulation of exponential distributions.  Applying Monte Carlo to solve the reduced integration problem leads to lower variance.
    
    Our new algorithm allows us to render multiple scattering on Smith microfacets with less variance than prior work, and, in the case of conductors, to do so without any bias. Additionally, our algorithm can also be used to accelerate the rendering of BSDFs containing volumetrically scattering layers, at reduced variance compared to standard Monte Carlo integration.

\ccsdesc[500]{Computing methodologies~Reflectance modeling}

\printccsdesc   
\end{abstract}  

\section{Introduction}

Volumetric light transport simulation in homogeneous slabs is a common problem in graphics and related fields.  Applications include paper \cite{Papas:2014:Paper}, skin layers \cite{Farrell:1992:Diffusion,Donner:2005:Light}, the rings of Saturn \cite{Blinn:1982:Light}, shielding materials \cite{Dwivedi:1982:New}, and volumetric models of scattering from rough surfaces \cite{Heitz:2016:Multiple}.

The most common application of slab transport in rendering is 
at a scale where the illumination can be considered laterally uniform.  The primary quantity of interest in this case is the bidirectional scattering distribution function (BSDF) for the slab, where the incoming and outgoing directions are given \cite{Hanrahan:1993:Reflection}.  
The BSDF follows from integrating over all possible paths inside the slab that connect these two directions, regardless of lateral displacement.  Despite the plane symmetry of the problem, very few closed-form solutions are known, requiring numerical techniques in the general case.

Both deterministic and stochastic methods can be used to evaluate slab BSDFs.  Methods such as discrete ordinates and doubling \cite{stam01,Jakob:2014:Comprehensive} are broadly applicable, but can require large precomputations and are not practical for rendering surfaces where multiple parameters vary spatially.   In contrast, Monte Carlo BSDF evaluation can be used in the case of general slabs and layered materials with no precomputation \cite{Guo:2018:Positionfree}, including full variation of all parameters over a surface.  However, this approach creates unwanted statistical noise and is expensive for thick highly-scattering materials.

The primary motivation of our work is to improve the efficiency of the Monte Carlo approach for slab BSDF evaluation for the special case of describing multiple scattering from rough surfaces.  The Smith microfacet model is widely used throughout computer graphics and including a multiple-scattering component is important to avoid unwanted darkening of rough materials.  
Due to the asymmetric phase function of Smith volumes, the Monte Carlo approach is the only known unbiased approach for evaluating this multiple scattering \cite{Heitz:2016:Multiple}.  Our primary contribution is a new efficient way to evaluate this multiple scattering.  We also demonstrate applications of the theory outside of rough surface scattering.

We reduce the variance and improve the efficiency of Monte Carlo slab BSDF evaluation by reducing the dimension of the path integral.  In previous work, the set of transport paths inside the slab are an alternating sequence of displacements and direction changes (\autoref{fig:slab-setup}).  We will show that the displacements can be preintegrated analytically, leaving only the directional dimensions of the path integral for Monte Carlo.  This extreme expected-value optimization reduces the dimension of the integration problem significantly and allows faster computation of full unbiased Monte Carlo solutions.  Because all three spatial coordinates of the path integral are integrated out, the resulting method is truly \emph{position free}, unlike previous work that evaluates slab BSDFs by sampling spatially 1D random walks in the slab \cite{Heitz:2016:Multiple,Guo:2018:Positionfree}.  Because the resulting state space depends only on directions, reasoning about bidirectional estimators becomes much simpler, in particular for the asymmetric phase functions appearing on microfacet surfaces~\cite{Heitz:2016:Multiple}.

After reviewing previous work in the next section, we define the problem in \cref{sec:3} and derive the position-free path space for homogeneous slabs in \cref{sec:slab}.  We apply our position-free estimators to a number of practical problems in computer graphics, such as rendering thin slabs (\cref{sec:slab-application}) and BSDFs with multiple scattering between the microfacets (\cref{sec:microfacet}).

\section{Related Work}\label{sec:previous}

Simulating the linear transport of particles in a homogeneous slab was one of the earliest applications of the Monte Carlo method \cite{Kahn:1949:Stochastic,goertzel1950proposed,Plass:1968:Monte}.  These works considered the plane-parallel integral equations of transfer where the phase space has a spatial component that is reduced from 3D to 1D (the depth within the slab).  These equations, which date back to the very beginning of radiative transfer, were solved using an analog Monte Carlo estimator that effectively performs a 1D random walk along the depth axis of the slab, with an appropriately modified extinction coefficient.  Estimation of radiance at the boundary (required for BSDF evaluation) was performed using next-event estimation (NEE) \cite{Kahn:1949:Stochastic}.  This approach has found a number of uses in graphics for evaluating BSDFs.  Hanrahan and Krueger~\cite{Hanrahan:1993:Reflection} rendered surfaces using discretized BSDFs precomputed in this way.  For rendering rough surfaces, Heitz et al.~\cite{Heitz:2016:Multiple,Dupuy:2016:Additional} applied the unbiased BSDF estimator with NEE directly at render time together with a blend of forward and adjoint estimates to reduce variance.  This approach was later extended to layered materials with full bidirectional generalizations and more advanced NEE that exploit the reduction to 1D in order to reduce variance \cite{Guo:2018:Positionfree,xia2020gaussian,gamboa2020efficient}.

\subsection{Position-free Monte Carlo}
Shortly after the introduction of these methods, it was recognized that homogeneous problems with plane symmetry could be simplified to fully position-free integration problems by noting that the Neumann series of collision densities in the medium is exactly described by 1D hyperexponential distributions, which can be efficiently computed using simple recurrence relations.  These derivations can be viewed as an expected-value transform of the previously mentioned spatially-1D Monte Carlo estimators, by analytically pre-integrating over the distances between collisions.  For the case of an infinite medium, these relations were derived in a number of ways \cite{Berger:1956:Reflection,Amster:1960:Euripus,Drawbaugh:1961:Solution,Chilton:1969:Imaging}.  The approach was applied to several problems by predetermining a fixed truncation of the Neumann series ahead of time based on known properties of the volume and then integrating over the directional dimensions of the reduced integral equation using Monte Carlo.
The extension of this approach to handle a slab was alluded to by Amster and Talley \cite{Amster:1964:Spatial} and given explicitly by Sears, who used it to derive the double-scattered BRDF for a half space with isotropic scattering \cite[(7.4.1)]{Sears:1975:Slow}.  

Our formalism is closely related to these works, but we go further to present a complete position-free derivation for rendering, including practical strategies for dealing with numerical issues, unbiased termination of the Neumann series using Russian roulette, and derivation of bidirectional estimators in the collapsed phase space. Unlike previous semi-position-free approaches in graphics, our formulation eliminates positions completely and achieves purely directional integration for lower variance. To the best of our knowledge, we are the first to apply these position-free Monte Carlo methods to rendering.

Our work is also closely related to the work of Wang et al.~\cite{Wang:2021:Position}, who present an approximate position-free formulation for Smith microsurfaces. Instead of analytically integrating positions, they replace position-dependent terms with the monostatic Smith shadowing function. While both our and their approach have similar efficiency, the approach of Wang et al.\ introduces significant bias at higher roughnesses. On dielectric microsurfaces, our analytic integration is not possible for paths that refract through the surface, and we selectively combine our approach with that of Wang et al.\ for such paths. This leads to a method with similar efficiency as that of Wang et al., but reduced bias.

\section{Background}\label{sec:3}

\begin{figure}[t]
    \centering
    \includegraphics[width=0.8\columnwidth]{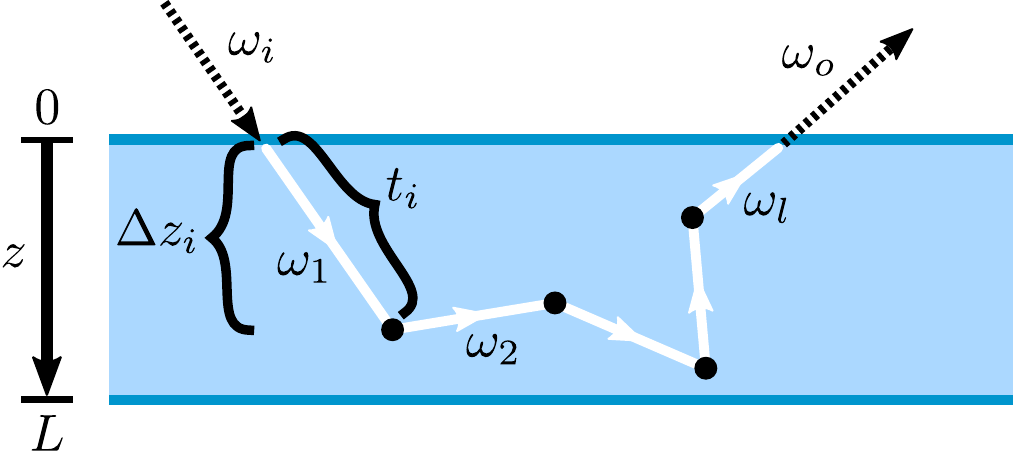}
    \caption{We show a sample path of length $l=5$ (counting the number of propagation distances) taken by a photon through a slab. The photon enters from direction $\DirAt{1}=\Dir_i$ and travels distance $t_1$ before scattering into direction $\DirAt{2}$. This process repeats until the path length $l$ is reached, at which point the photon leaves in prescribed direction $\DirAt{l}=\DirO$. \label{fig:slab-setup}}\vspace{-3mm}
\end{figure}

In this paper, we consider homogeneous media contained in a \emph{slab} of infinite $x$-$y$ extent. The slab occupies the $z$-span $[0, \SlabD]$, where the $z$ axis points down into the slab (see \cref{fig:slab-setup}).

Associated with the medium is the phase function $\Phase(\DirX, \DirY)$ that describes the density with which photons travelling in direction $\DirX$ scatter toward direction $\DirY$ after a collision.  The phase function may integrate to less than one to incorporate absorption.

The medium has homogeneous extinction coefficient $\ExtCoeff$, which gives rise to the transmittance
\begin{align}
    \Trans(t, \ExtCoeff) = \text{e}^{-t \ExtCoeff}
\end{align}
and free-flight probability density function (PDF)
\begin{align}
    \FFPdf{t, \ExtCoeff} = \ExtCoeff \text{e}^{-t \ExtCoeff}.
\end{align}

We are particularly interested in the density of photons that enter the slab from direction $\DirI$ and leave in direction $\DirO$. We can formulate this problem as the integral over all possible paths a photon could take through the slab. We show one such path of length $l$ (counting the number of propagation distances) in \autoref{fig:slab-setup}: The photon enters from direction $\DirAt{1} = \DirI$, travels a distance $\DistAt{1}$ with PDF $\FFPdf{\DistAt{1}, \ExtCoeff}$, before scattering into a new direction $\DirAt{2}$. This process continues until it reaches length $l$, at which point it leaves the slab in prescribed direction $\DirAt{l} = \DirO$. Without loss of generality, we assume the photon enters from the top at $z=0$; if not, we can simply negate $\DirI$ and $\DirO$.

The total contribution of all paths of length $l$ is
\begin{align}
    I_l &= \int_\PathSpace \left(\prod_{i=1}^{l-1} \Phase(\DirAt{i}, \DirAt{i+1})\right) \left( \prod_{i=1}^{l-1} \FFPdf{\DistAt{i}, \ExtCoeff} \right) \Trans(\DistAt{l}, \ExtCoeff) \mathrm{d}\mu(\bar{t},\bar{\Dir}), \label{eqn:measurement}
\end{align}
where $\PathSpace$ is the space of paths that lie within the slab, and $\mathrm{d}\mu(\bar{t},\bar{\Dir})$ = $\mathrm{d}\mu(\bar{t}) \times \mathrm{d}\mu(\bar{\Dir})$ is the product of standard Lebesgue measures of distances $\bar{t}=t_1\ldots t_l$ and solid angle measures of directions $\bar{\Dir}=\Dir_1\ldots \Dir_l$. The quantity we seek to estimate is the sum $\sum_{l=2}^\infty I_l$ over all lengths.

The integrand in \cref{eqn:measurement} consists of three components: The directional density $\prod \Phase(\DirAt{i}, \DirAt{i+1})$ of the path; the free-flight density $\prod \FFPdf{\DistAt{i}, \ExtCoeff}$ of the travel distances $\DistAt{1} \ldots \DistAt{l-1}$; and the probability $\Trans(\DistAt{l})$ of the photon exiting the slab from the last path vertex. Note that directions $\DirAt{1} = \DirI, \DirAt{l}=\DirO$ and distance $\DistAt{l}$ are not free variables; the latter of which is determined by the distance of the last path vertex to the boundary.

\section{Position-free transport in a slab}\label{sec:slab}

The goal of this section is to show that all distance dimensions $\DistAt{i}$ in \cref{eqn:measurement} can be integrated analytically. We begin by noting that the directional terms do not depend on distances, and we can move part of the integration inside:
\begin{align}
    I_l &= \int_\DirSpace \left(\prod_{i=2}^{l-1} \Phase(\DirAt{i}, \DirAt{i+1})\right) \underbrace{\int_\DistSpace \left( \prod_{i=1}^{l-1} \FFPdf{\DistAt{i}, \ExtCoeff} \right) \Trans(\DistAt{l}) \mathrm{d}\mu(\bar{t})}_{\ExitP(\Dir_1,\ldots,\Dir_l)} \mathrm{d}\mu(\bar{\Dir})
\end{align}
where $\DirSpace$ and $\DistSpace$ are the spaces of directions and distances, respectively (with $\PathSpace = \DirSpace \times \DistSpace$).
The inner integration is the probability $\ExitP(\Dir_0,\ldots,\Dir_l)$ of a photon exiting the slab, conditioned on the directions it takes after each collision. We will now show how this probability can be derived in closed form.

\subsection{Simplifying to 1D}

Our first step is to simplify the integration problem. We will perform a change in variables from the distance $\DistAt{i}$ the photon travelled, to the (absolute) \emph{height difference} between collisions $\DeltaZAt{i} = \DistAt{i}  \cdot \Abs{\ZComponent{\DirAt{i}}}$, where $\ZComponent{\Dir}$ refers to the $z$-component of vector $\Dir$.

This type of simplification is common in slab problems, having been used extensively outside~\cite{Amster:1960:Euripus} and within graphics~\cite{Hanrahan:1993:Reflection,Heitz:2016:Multiple,Dupuy:2016:Additional,Guo:2018:Positionfree,xia2020gaussian,gamboa2020efficient}. We briefly review it here before deriving our position-free formulation. After the change of variables, we obtain the relation
\begin{align}
    \FFPdf{\DistAt{i}, \ExtCoeff} \mathrm{d} \DistAt{i}
        &= \frac{1}{\Abs{\ZComponent{\DirAt{i}}}} \FFPdf{\frac{\DeltaZAt{i}}{\Abs{\ZComponent{\DirAt{i}}}},\ExtCoeff} \mathrm{d} \DeltaZAt{i} \\
        &= \FFPdf{\DeltaZAt{i}, \ExtCoeffAt{i}} \mathrm{d} \DeltaZAt{i} \nonumber \\
        \quad \text{with} \quad \ExtCoeffAt{i} &= \frac{\ExtCoeff}{\Abs{\ZComponent{\DirAt{i}}}}. \label{eqn:angle-dependent-sigma}
\end{align}
and similarly
\begin{align}
    \Trans(\DistAt{i}, \ExtCoeff)  = \Trans(\DeltaZAt{i}, \ExtCoeffAt{i}).
\end{align}
These equations tell us that collisions in the 3D slab are equivalent to collisions in a ``densified'' 1D medium, where the extinction coefficient increases as photon directions become less vertical. This formulation allows for integrating explicitly over the depths of collisions, rather than the distances travelled between them.

With the projection to 1D, we can reformulate the problem of computing $\ExitP$ to a simpler one: Given a \emph{height distribution} $\HeightPdf{l-1}(z)$, which represents the probability density that the photon will collide at height $z$ after travelling $l-1$ distances in the medium, we can express the exit probability of the photon as
\begin{align}
    \ExitP(\Dir_{1,\ldots,l}) = \begin{cases}
        \mathlarger{\ExitUp = \int_0^\SlabD \HeightPdf{l-1}(z) \Trans(z, \ExtCoeffAt{l}) \mathrm{d}z} &\text{if $\ZComponent{\Dir_l} < 0$} \\
        \mathlarger{\ExitDown = \int_0^\SlabD \HeightPdf{l-1}(z) \Trans(L - z, \ExtCoeffAt{l}) \mathrm{d}z} &\text{else.} \label{eqn:exit-probability}
    \end{cases}
\end{align}
This is simply the PDF of colliding at $z$ after $l-1$ propagations, multiplied by the probability of exiting the slab from $z$. We differentiate between the case of the photon exiting the top ($z=0$) and bottom ($z=\SlabD$) interface.

From here, we will proceed as follows: First, we show how the height distribution can be derived. Second, we show that it can be represented in closed form as a hyperexponential distribution (a sum of exponentials). Finally, we show that given such a height distribution, \cref{eqn:exit-probability} can be computed in closed form.

\subsection{Distribution of Heights}

We begin with the straightforward case of $\HeightPdf{1}(z)$, which is the probability density of a photon colliding at $z$ after propagating once in the medium. This is equivalent to the free-flight PDF \cite[p.5]{Kahn:1949:Stochastic}
\begin{align}
    \HeightPdf{1}(z) &= \FFPdf{z,\ExtCoeffAt{1}}.
\end{align}
What about the distribution after travelling $i$ distances? This can be written recursively in terms of the density $\HeightPdf{i-1}$ and depends on the direction of travel since the last collision: If the photon is moving down into the slab, then its collision density is equivalent to the density of colliding at a point $y<z$ above $z$, and then colliding again at $z$:
\begin{align}
    h_i^\downarrow(z) &= \int_0^z h_{i-1}(y) \FFPdf{z - y,\sigma_i} \mathrm{d} y. \label{eqn:height-dist-down}
\end{align}
If the photon is travelling upward, then its collision density is equivalent to colliding at some point $y>z$ below $z$, then colliding again at $z$:
\begin{align}
    h_i^\uparrow(z) &= \int_z^L h_{i-1}(y) \FFPdf{y - z,\sigma_i} \mathrm{d} y. \label{eqn:height-dist-up}
\end{align}

\subsection{Height Distribution in Closed Form} \label{sec:height-distribution}

With the height distribution defined, we now turn to solving it in closed form. We will claim that $h_i(z)$ is a sum of exponentials,
\begin{align}
    h_i(z) &= \sum_{j=1}^{N_i} \seqA{i}{j} \Trans(z, \seqB{i}{j}) \label{eqn:height-distribution-base}
\end{align}
for some $N_i,\seqA{i}{j}, \seqB{i}{j}$. The proof follows by induction. The base case is trivial, with $\HeightPdf{1}(z) = \FFPdf{z,\ExtCoeffAt{1}}$ and $N_1=1$, $\seqA{1}{1} = \seqB{1}{1} = \ExtCoeffAt{1}$. For the induction step, we distinguish between two cases, depending on whether the photon is moving up or down.

\subsubsection{Downward moving photon} For the ``down'' case, we give a sketch of the proof below, and provide a full proof in the supplemental. We begin by expanding \cref{eqn:height-dist-down}:
\begin{align}
    h_{i+1}^\downarrow(z)
        =& \int_0^z h_i(y) \FFPdf{z - y,\sigma_{i+1}} \mathrm{d} y \\
        =& \sum_{j=1}^{N_i} \seqA{i}{j} \int_0^z \Trans(y, \seqB{i}{j}) \FFPdf{z - y,\sigma_{i+1}} \mathrm{d} y \\
        =& \sum_{j=1}^{N_i} \seqA{i}{j} \frac{\sigma_{i+1}}{\sigma_{i+1} - \seqB{i}{j}} \left(e^{-z \seqB{i}{j}} - e^{-z \sigma_{i+1}}\right) \\
        =& \left( \sum_{j=1}^{N_i} \frac{\seqA{i}{j} \: \sigma_{i+1}}{\sigma_{i+1} - \seqB{i}{j}} \Trans(z, \seqB{i}{j}) \right) - \left( \sum_{j=1}^{N_i} \frac{\seqA{i}{j} \: \sigma_{i+1}}{\sigma_{i+1} - \seqB{i}{j}} \right ) \Trans(z, \sigma_{i+1}) \nonumber \\
        =& \sum_{j=1}^{N_{i+1}} \seqA{i+1}{j} \Trans(z, \seqB{i+1}{j}).
\end{align}
If the height distribution is a sum of exponentials, then it remains a sum of exponentials after the photon travels an additional (downward) segment. The coefficients of the new distribution are
\begin{align}
    \seqA{i+1}{j}^\downarrow &= \begin{cases}
        \seqA{i}{j} \frac{\sigma_{i+1}}{\sigma_{i+1} - \seqB{i}{j}}
        &\quad \text{if $j<N_{i+1}$} \\
        \sum_{j=1}^{N_i} -\seqA{i+1}{j} &\quad \text{else}
    \end{cases} \label{eqn:height-dist-down-closed} \\
    \seqB{i+1}{j}^\downarrow &= \begin{cases}
        \seqB{i}{j} &\quad \text{if $j<N_{i+1}$} \\
        \sigma_{i + 1} &\quad \text{else.}
    \end{cases} \quad\text{and}\quad N_{i+1}^\downarrow = N_{i} + 1 \nonumber
\end{align}
\subsubsection{Upward moving photon} The ``up'' case is very similar to the down case, and we give a proof in the supplemental. The distribution remains a sum of exponentials, with coefficients
\begin{align}
    \seqA{i+1}{j}^\uparrow &= \begin{cases}
        \seqA{i}{j} \frac{\sigma_{i+1}}{\sigma_{i+1} + \seqB{i}{j}}
        &\quad \text{if $j<N_{i+1}$} \\
        \sum_{j=1}^{N_i} -\seqA{i+1}{j} \Trans(L, \ExtCoeffAt{i+1} + \seqB{i}{j}) &\quad \text{else}
    \end{cases} \label{eqn:height-dist-up-closed} \\
    \seqB{i+1}{j}^\uparrow &= \begin{cases}
        \seqB{i}{j} &\quad \text{if $j<N_{i+1}$} \\
        -\sigma_{i + 1} &\quad \text{else.}
    \end{cases} \quad\text{and}\quad N_{i+1}^\uparrow = N_{i} + 1 \nonumber
\end{align}

\subsection{Exit Probability} \label{sec:exit-prob-closed}

Finally, given the height distribution, we can now derive the exit probability by inserting \cref{eqn:height-distribution-base} into \cref{eqn:exit-probability}. The probability reduces to a simple sum of transmittances (see supplemental for derivation):
\begin{align}
    \ExitUp(\Dir_0,\ldots,\Dir_l) &= \sum_{j=1}^{N_i} \frac{\seqA{i}{j}}{\ExtCoeffAt{l} + \seqB{i}{j}} \left( 1 - \Trans(L, \ExtCoeffAt{l} + \seqB{i}{j}) \right) \label{eqn:exitp-up} \\
    \ExitDown(\Dir_0,\ldots,\Dir_l) &= \sum_{j=1}^{N_i} \frac{\seqA{i}{j}}{\ExtCoeffAt{l} - \seqB{i}{j}} \left( \Trans(L, \seqB{i}{j}) - \Trans(L, \ExtCoeffAt{l}) \right) \label{eqn:exitp-down}
\end{align}

\subsection{Semi-infinite slab}\label{sec:halfspace}

A useful special case arises for a semi-infinite slab, i.e. $\SlabD \rightarrow \infty$. In this case, the ``up'' case simplifies to
\begin{align}
    N_{i+1}^\uparrow &= N_{i}, \quad \seqB{i+1}{j}^\uparrow = \seqB{i}{j} \quad\text{and}\quad \seqA{i+1}{j}^\uparrow = \seqA{i}{j} \frac{\sigma_{i+1}}{\seqB{i}{j} + \sigma_{i+1}},  \label{eqn:height-dist-up-infinite}
\end{align}
where the number of exponentials stays unchanged, and only the amplitude of the existing exponentials is rescaled. The ``down'' case remains identical to the finite slab.

\begin{algorithm}[t]
	\DontPrintSemicolon
    \Class{\emph{HeightDistribution}}{
        \Def{$N, a[], b[]$\;}
        \Fn{$\mathrm{addBounce}(\Dir, \ExtCoeff, \SlabD)$}{
            \If(\tcp*[h]{Base case}){$N = 0$}{
                $a[1] \leftarrow \ExtCoeff$\;
                $b[1] \leftarrow \ExtCoeff$\;
            }\ElseIf(\tcp*[h]{``Up'' case, \cref{eqn:height-dist-up-closed}}){$\ZComponent{\Dir} < 0$}{
                \For {$i \leftarrow 1\ldots N$}{
                    $a[i] \leftarrow a[i] \frac{\ExtCoeff}{\ExtCoeff + b[i]}$\;
                }
                $a[N + 1] \leftarrow \sum_{i=1}^{N} -a[i] \Trans(L, \ExtCoeff + b[i])$\;
                $b[N + 1] \leftarrow -\ExtCoeff$\;
            }\Else(\tcp*[h]{``Down'' case, \cref{eqn:height-dist-down-closed}}){
                \For {$i \leftarrow 1\ldots N$}{
                    $a[i] \leftarrow a[i] \frac{\ExtCoeff}{\ExtCoeff - b[i]}$\;
                }
                $a[N + 1] \leftarrow \sum_{i=1}^{N} -a[i]$\;
                $b[N + 1] \leftarrow \ExtCoeff$\;
            }
            $N \leftarrow N + 1$\;
        }
        \Fn{$\ExitP(\Dir, \ExtCoeff, \SlabD)$}{
            \If(\tcp*[h]{``Up'' case}){$\ZComponent{\Dir} < 0$}{
                \Return{$\sum_{i=1}^{N} \frac{a[i]}{\ExtCoeff + b[i]} \left( 1 - \Trans(L, \ExtCoeff + b[i]) \right)$}\tcp*{\cref{eqn:exitp-up}}
            }\Else(\tcp*[h]{``Down'' case}){
                \Return{$\sum_{i=1}^{N} \frac{a[i]}{\ExtCoeff - b[i]} \left( \Trans(L, b[i]) - \Trans(L, \ExtCoeff) \right)$} \tcp*{\cref{eqn:exitp-down}}
            }
        }
    }
    \caption{Implementation of the closed-form height distribution (\cref{sec:height-distribution}) and computation of $\ExitP$ (\cref{sec:exit-prob-closed}).}\label{alo:height-dist}
\end{algorithm}

The exit probabilities simplify as well, with
\begin{align}
    \ExitUp(\Dir_1,\ldots,\Dir_l) &= \sum_{j=1}^{N_i} \frac{\seqA{i}{j}}{\seqB{i}{j} + \ExtCoeffAt{l}}  \\
    \ExitDown(\Dir_1,\ldots,\Dir_l) &= 0
\end{align}

\subsection{Discussion}\label{sec:discussion}

We have derived closed form expressions for the height distribution of a photon after an arbitrary number of collisions in a homogeneous slab, as well as its probability of exiting the slab. The reflectance of a slab can then be computed in a position-free way by sampling the directions of the path, and then evaluating $\ExitP$ in closed form.

To compute $\ExitP$ given a set of directions $\Dir_1,\ldots,\Dir_l$, we first require computing the height distribution $h_l$.  The coefficients of the distribution are initialized with the base case, and we run the update rules \cref{eqn:height-dist-down-closed} or \cref{eqn:height-dist-up-closed} for each direction in sequence. Finally, we compute the exit probability with \cref{eqn:exitp-up} or \cref{eqn:exitp-down} to obtain the final result.

We give pseudo-code for computing $h_l$ and $\ExitP$ in \cref{alo:height-dist}. Doing so requires storing up to $2l$ coefficients, $\seqA{i}{j}$ and $\seqB{i}{j}$, and requires $O(l^2)$ total operations. For the semi-infinite case, the operations involve only simple arithmetic; for the finite slab, we require evaluating exponentials as well. Using these building blocks, we will now derive position-free integration algorithms for several practical applications in graphics.

\section{Application: Slab BSDF} \label{sec:slab-application}

A first obvious application of our method is for BSDFs that model far-field scattering by slabs of homogeneous media. These slabs could act as a component in specialized BSDFs such as those for paper~\cite{Papas:2014:Paper}, or in more general layered material frameworks either enclosed by dielectric interfaces~\cite{Guo:2018:Positionfree,weidlich2007arbitrarily} or in a stack of slabs with index-matched interfaces~\cite{Wang:2021:Spongecake}. In the following, we will focus on the straightforward case of a BSDF representing reflectance from a single, index-matched homogeneous slab. More sophisticated applications can then be easily built using this basic case as a building block.

\begin{algorithm}[t]
	\DontPrintSemicolon
    \Fn{$\mathrm{analogSlab}(l, \DirI, \DirO, \SlabD)$}{
        $z \leftarrow 0$\;
		$\Dir = \DirI$\;
		$\mathrm{result} \leftarrow 0$\;
        \For {$i \leftarrow 1\ldots l-1$}{
            $z \leftarrow z + \mathrm{sampleFreeFlight}(\sigma(\Dir))$\;
            \If{$z \not \in [0, \SlabD]$}{\Break}
            $z_\mathrm{exit}\leftarrow \ZComponent{\DirO} < 0 \;?\; z : L - z$\;
            $\mathrm{result} \leftarrow \mathrm{result} + \Phase(\Dir, \DirO) \cdot \Trans(z_\mathrm{exit}, \ExtCoeff(\DirO))$\;
            $\Dir \leftarrow \mathrm{scatter}(\Dir)$\;
        }
        \Return{$\mathrm{result}$}
    }
    \caption{Baseline \emph{analog} estimation of reflectance from a slab.}\label{alo:analog-unidir}
\end{algorithm}
\begin{algorithm}[t]
	\DontPrintSemicolon
    \Fn{$\mathrm{positionFreeSlab}(l, \DirI, \DirO, \SlabD)$}{
		$\ClassName{HeightDistribution}\; h$\;
		$\Dir = \DirI$\;
		$\mathrm{result} \leftarrow 0$\;
        \For {$i \leftarrow 1\ldots l-1$}{
            $h$.addBounce($\Dir, \ExtCoeff(\Dir), \SlabD$)\;
            $\mathrm{result} \leftarrow \mathrm{result} + \Phase(\Dir, \DirO) \cdot h.\ExitP(\DirO, \ExtCoeff(\DirO), \SlabD)$\;
            $\Dir \leftarrow \mathrm{scatter}(\Dir)$\;
        }
        \Return{$\mathrm{result}$}
    }
    \caption{Position-free estimation of reflectance from a slab.}\label{alo:position-free-unidir}
\end{algorithm}

\subsection{Algorithm}

Transport on the slab can be readily estimated using Monte Carlo by simulating consecutive events (collisions and scattering) that occur on the photon path. We show one such baseline \emph{analog} simulator in \cref{alo:analog-unidir}, computing contributions of paths up to length $l$. This algorithm is equivalent to the unidirectional estimator of Guo et al.~\cite{Guo:2018:Positionfree}. We are given routines for sampling directions and propagation distances, and we accumulate the amount of energy leaving the slab (using next-event estimation~\cite{Pharr:2016:Physically}). We are also given a function $\ExtCoeff(\Dir)$ that computes the 1D extinction coefficient, which here is simply $\ExtCoeff(\Dir) = \ExtCoeff \cdot \Abs{\ZComponent{\Dir}}$; in \cref{sec:microfacet}, we will substitute a different $\ExtCoeff(\Dir)$ to simulate microfacets.

We can readily modify this baseline estimator to perform fully position-free integration instead. We show pseudo-code of such an estimator in \cref{alo:position-free-unidir}. We replace the tracking and sampling of explicit positions or heights with the computation of closed-form height distributions $h_l$. Next-event estimation can then be performed using the closed-form probability $\ExitP$ of leaving the slab, instead of the transmittance as in \cref{alo:analog-unidir}.

\subsubsection{Numerical Stability} One potential issue with \cref{alo:height-dist,alo:position-free-unidir} lies in numerical stability. Computing $\ExitP$ relies on taking a sum of ratios with potentially different signs and large variations in magnitude. Additionally, parts of the algorithm compute fractions with factors of $1/(\ExtCoeff - b[i])$. This raises concerns about numerical stability for long paths (as $N$ becomes large) and cases when $\ExtCoeff$ approaches $b[i]$, e.g.\ if cosines $(\Dir_i)_z = (\Dir_j)_z$ are very close for some pair of directions in the path. While we did not find it feasible to completely \emph{prevent} these numerical issues completely, we can easily \emph{detect} them. Past a certain number of bounces (10 in our implementation), or if ratios become unstable, we fall back to analog simulation of the path. This still gives an unbiased result, at the cost of variance. We found that only an insubstantial fraction of paths cause numerical concerns ($\approx$ 0.1\%), and the overall efficiency of our algorithm remains substantially better than purely analog simulation.  

These numerical issues were also noted in earlier works, and mitigation strategies to reduce the equal-cosine case to gamma distributions and/or clamping heuristics have been proposed \cite{Berger:1956:Reflection,Amster:1960:Euripus,Amster:1964:Spatial}.  Although we do not use the clamping strategy currently, it shows promising results, and poses an interesting avenue for future work.

\subsection{Results}

We implemented our method in an open source offline rendering framework~\cite{Bitterli:2018:Tungsten} as a material model representing homogeneous slabs, and in a simple C++ testbed for benchmarking BSDF evaluations. We compare our position-free framework to the approach of Guo et al.~\cite{Guo:2018:Positionfree}, which simulates explicit depths within the slab. For the method of Guo et al., we tried both bidirectional and unidirectional simulation, and only show the more efficient of the two (unidirectional in this case).

\begin{figure*}[t]
	\centering
	\setlength{\tabcolsep}{0pt}
	\hspace{-6mm}
	\begin{tabular}{cc@{\hspace{11mm}}cc}
	    \cmidrule(lr{11mm}){1-2}\cmidrule(lr){3-4}\\[-5mm]\\
		\parbox{0.24\textwidth}{\input{gfx/slab-plots-new/SlabReflectance-thetai-0-L-0.5-g--0.50.pgf}}  & \parbox{0.24\textwidth}{\input{gfx/slab-plots-new/SlabTTUV-thetai-0-L-0.5-g--0.50.pgf}}  & \parbox{0.24\textwidth}{\input{gfx/slab-plots-new/SlabReflectance-thetai-0-L-1.0-g-0.50.pgf}}  & \parbox{0.24\textwidth}{\input{gfx/slab-plots-new/SlabTTUV-thetai-0-L-1.0-g-0.50.pgf}} \\
		\parbox{0.24\textwidth}{\input{gfx/slab-plots-new/SlabReflectance-thetai-45-L-2.0-g--0.50.pgf}} & \parbox{0.24\textwidth}{\input{gfx/slab-plots-new/SlabTTUV-thetai-45-L-2.0-g--0.50.pgf}} & \parbox{0.24\textwidth}{\input{gfx/slab-plots-new/SlabReflectance-thetai-45-L-8.0-g-0.00.pgf}}  & \parbox{0.24\textwidth}{\input{gfx/slab-plots-new/SlabTTUV-thetai-45-L-8.0-g-0.00.pgf}} \\
		\parbox{0.24\textwidth}{\input{gfx/slab-plots-new/SlabReflectance-thetai-80-L-0.5-g--0.50.pgf}} & \parbox{0.24\textwidth}{\input{gfx/slab-plots-new/SlabTTUV-thetai-80-L-0.5-g--0.50.pgf}} & \parbox{0.24\textwidth}{\input{gfx/slab-plots-new/SlabReflectance-thetai-80-L-1.0-g-0.50.pgf}} & \parbox{0.24\textwidth}{\input{gfx/slab-plots-new/SlabTTUV-thetai-80-L-1.0-g-0.50.pgf}} \\
		\parbox{0.24\textwidth}{\input{gfx/slab-plots-new/SlabReflectance-thetai-80-L-4.0-g--0.90.pgf}} & \parbox{0.24\textwidth}{\input{gfx/slab-plots-new/SlabTTUV-thetai-80-L-4.0-g--0.90.pgf}} & \parbox{0.24\textwidth}{\input{gfx/slab-plots-new/SlabReflectance-thetai-0-L-1.0-g-0.90.pgf}}  & \parbox{0.24\textwidth}{\input{gfx/slab-plots-new/SlabTTUV-thetai-0-L-1.0-g-0.90.pgf}}
	\end{tabular}
	\includegraphics[width=0.25\textwidth]{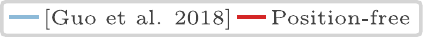}
    \caption{We show mean BSDF value (left half of columns) and inverse efficiency (right half of columns, lower is better) for reflectance from a homogeneous slab with parameters (thickness, mean phase cosine and incident angle) randomly selected from a list. Our position-free estimator (red curve) outperforms the baseline estimator of Guo et al.~\cite{Guo:2018:Positionfree} (blue curve) over nearly all parameters and angles, usually significantly so. The mean reflectance values computed by both estimators match perfectly.\label{fig:slab-efficiency}}
\end{figure*}

We can measure estimator performance directly by comparing the \emph{inverse efficiency}, i.e.\ the average cost of evaluating the estimator times its variance. An estimator performs better if its inverse efficiency is lower than that of other estimators. In \cref{fig:slab-efficiency}, we show a comprehensive evaluation of our estimator compared to the baseline estimator of Guo et al. over randomly selected slab thicknesses, incident angles and phase mean cosines.
We plot the expected value of BSDF evaluations produced by both methods (which match exactly), as well as the inverse efficiency of both estimators over different inclinations of $\DirO$. Our proposed estimator significantly outperforms the baseline in almost all cases, even when comparing the cost of pure BSDF evaluation. In real rendering scenarios, the cost of BSDF evaluation is small compared to path tracing, and the reduction in variance at equal render time becomes even more significant. 

\begin{figure}[t]
    \newcommand{\insetratio}{0.35}%
    \setlength{\bigw}{0.49\columnwidth}%
    \setlength{\insetw}{\insetratio\bigw}%
    \setlength{\gridw}{0.5\columnwidth}%
    \begin{tikzpicture}[x=\gridw,y=\gridw,every text node part/.style={align=center}]
        \node[anchor=north west] at (0, 0) {\includegraphics[width=\bigw]{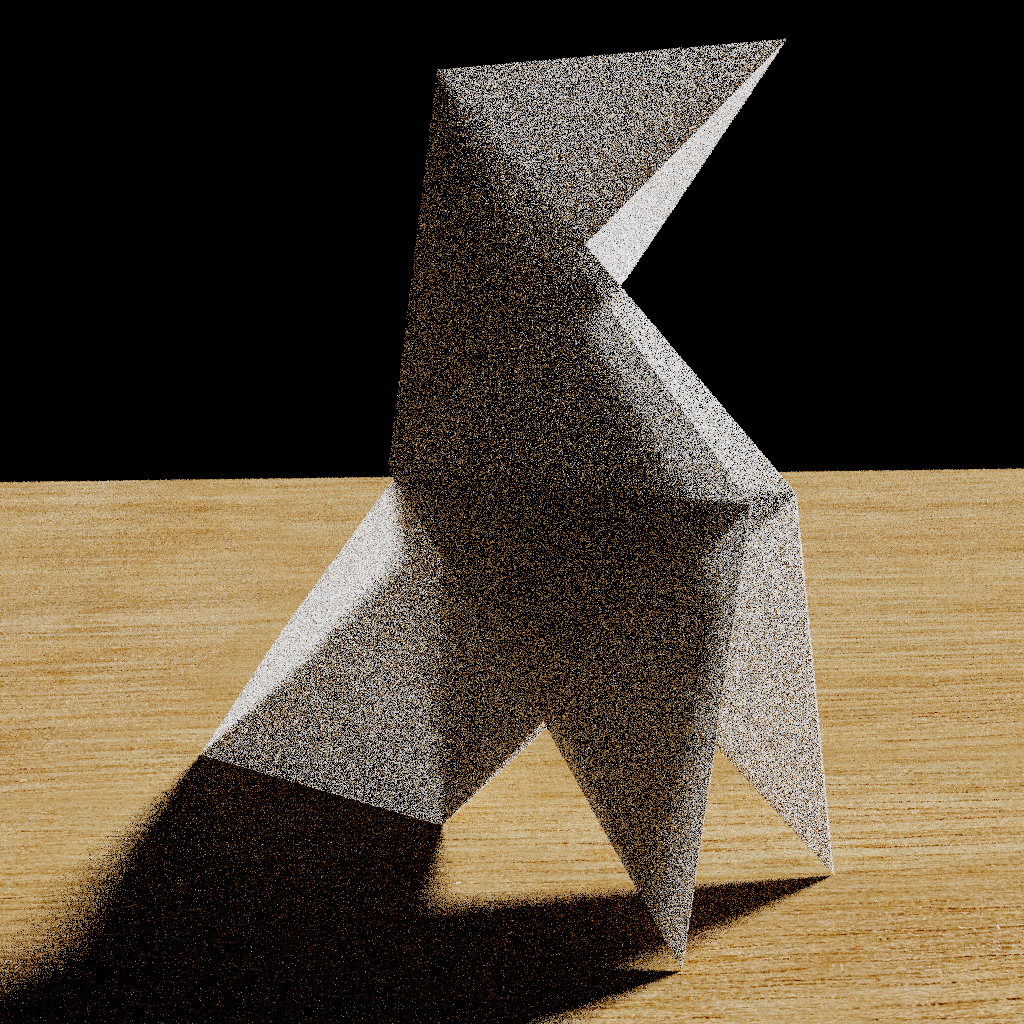}};
        \node[anchor=north west] at (1, 0) {\includegraphics[width=\bigw]{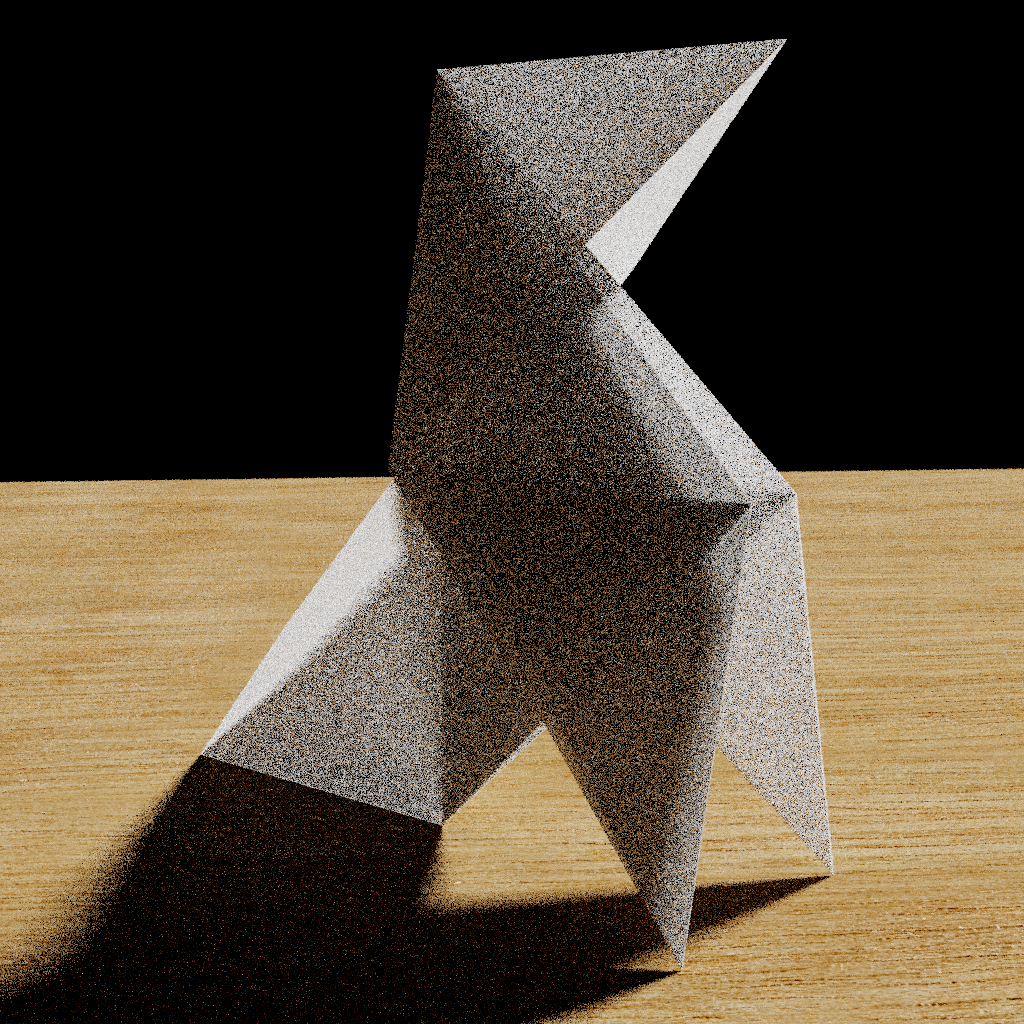}};
        \draw[draw=colorA,line width=1.5] (0+314/1024, -521/1024) rectangle ++(120/1024, -120/1024);
        \draw[draw=colorA,line width=1.5] (1+314/1024, -521/1024) rectangle ++(120/1024, -120/1024);
        \node[anchor=north west] at (0, 0.0) {\includegraphics[width=\insetw]{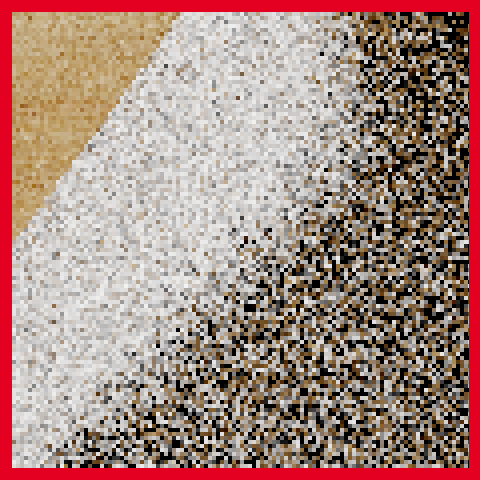}};
        \node[anchor=north west] at (1, 0.0) {\includegraphics[width=\insetw]{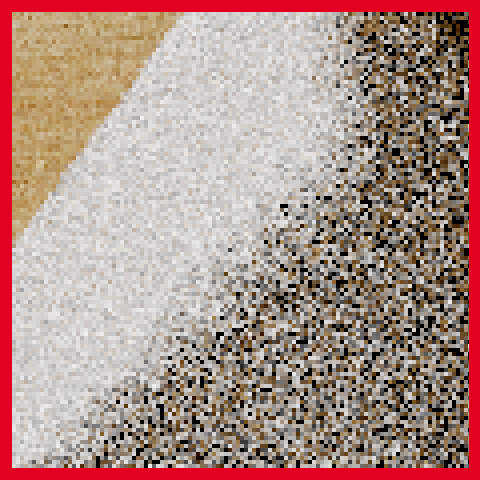}};
        \node[anchor=north] at (0+0.54*\insetratio, -0.02) {\footnotesize\bfseries\contour{black}{\color{white}MSE: 0.0049}};
        \node[anchor=north] at (1+0.54*\insetratio, -0.02) {\footnotesize\bfseries\contour{black}{\color{white}MSE: 0.0030}};
        \node[anchor=north east] at (1-0.02, -0.02) {\bfseries\contour{black}{\color{white}\citeauthor{Guo:2018:Positionfree}}};
        \node[anchor=north east] at (1-0.02, -0.02 - 1.5em) {\small\bfseries\contour{black}{\color{white}714ms}};
        \node[anchor=north east] at (2-0.02, -0.02) {\bfseries\contour{black}{\color{white}Ours}};
        \node[anchor=north east] at (2-0.02, -0.02 - 1.5em) {\small\bfseries\contour{black}{\color{white}756ms}};
    \end{tikzpicture}\\[-2mm]%
    \caption{We render a papercraft figure using a BSDF computing reflectance from a homogeneous slab. The slab parameters are $\SlabD=2.5$ and $\ExtCoeff=1$, and we use a Henyey-Greenstein phase function with mean cosine $g=-0.5$. Our position-free integrator (right, \cref{alo:position-free-unidir}) significantly outperforms the analog baseline (left, \cref{alo:analog-unidir}) at roughly equal render times. \label{fig:origami-figure}}
\end{figure}

To show a practical example, we show a simple origami figure rendered with a homogeneous slab material in \cref{fig:origami-figure}, produced by both algorithms. At equal render time, our algorithm provides significantly reduced variance, both in terms of visible noise and MSE. Pre-integrating parts of the integration problem, and thus reducing its dimensionality, directly reduces the variance of the Monte Carlo estimator for the slab.

\section{Application: Multiple Scattering on Smith Microfacets} \label{sec:microfacet}

The main application of our theory is to multiple scattering for microfacet BSDFs~\cite{Walter:2007:Microfacet}. Heitz et al.\cite{Heitz:2016:Multiple} propose computing multiple scattering on microfacet surfaces under the Smith model by formulating it as a random walk on the stochastic heightfield. Dupuy et al.\cite{Dupuy:2016:Additional} further simplify this approach by showing the problem is equivalent to transport in a homogeneous half space, with angle-dependent extinction coefficient
\begin{align}
    \ExtCoeff_\mathrm{smith}(\Dir) = \begin{cases}
        \Lambda(\Dir) \Abs{\ZComponent{\Dir}} &\quad\text{if $\ZComponent{\Dir} < 0$} \\
        (1 + \Lambda(-\Dir)) \Abs{\ZComponent{\Dir}} &\quad\text{else,} \label{eqn:smith-extinction}
    \end{cases}
\end{align}
where $\Lambda(\Dir)$ is the Smith lambda function~\cite{Walter:2007:Microfacet,Heitz:2016:Multiple}. Although this approach is an exact (albeit probabilistic) solution, the drawback is that every evaluation of the BSDF involves a random walk through the medium (or heightfield), which is costly and introduces variance.

\subsection{Algorithm}

We can directly apply our position-free path integral to the semi-infinite formulation of Dupuy et al. \cite{Dupuy:2016:Additional} and analytically integrate out the spatial dimensions of the random walk. We combine their extinction coefficient (\cref{eqn:smith-extinction}) with ours (\cref{eqn:angle-dependent-sigma}) to obtain
\begin{align}
    \ExtCoeff(\Dir) = \begin{cases}
        \Lambda(\Dir) &\quad\text{if $\ZComponent{\Dir} < 0$} \\
        1 + \Lambda(\Dir) &\quad\text{else.} \label{eqn:smith-extinction-ours}
    \end{cases}
\end{align}
Inserting \cref{eqn:smith-extinction-ours} into \cref{alo:position-free-unidir} immediately leads to a position-free estimator of multiple scattering on microfacet surfaces. While preferable to fully analog simulation, this algorithm would be inferior to the one proposed by Heitz et al.\cite{Heitz:2016:Multiple}. On microfacet surfaces, a significant source of noise comes from phase function sampling, as opposed to free-flight sampling. While our algorithm eliminates variance from the latter, it does nothing for the former. Heitz et al. \cite{Heitz:2016:Multiple} propose to start the random walk randomly either at $\DirI$ (``forward'') or $\DirO$ (``backward''), and to perform multiple importance sampling (MIS~\cite{Veach:1995:Optimally}) to combine evaluation from both directions. The same principle can be applied to our position-free algorithm (\cref{alo:position-free-unidir}) to obtain an estimator that consistently outperforms that of Heitz et al. \cite{Heitz:2016:Multiple}.

\subsubsection{Bidirectional Estimators} An unexpected benefit of the position-free formulation relates to the nature of the medium formulated by Dupuy et al.~\cite{Dupuy:2016:Additional}. Their extinction coefficient (\cref{eqn:smith-extinction}) not only depends on direction, but is also \emph{non-symmetric}: $\ExtCoeff_\mathrm{smith}(\Dir) \neq \ExtCoeff_\mathrm{smith}(-\Dir)$. Transmittance and the free-flight PDF thus are no longer equivalent if the travel direction is reversed; reciprocity holds only for complete paths that start and end on the boundary, which guarantees overall reciprocity of the BRDF. This means that reciprocity does \emph{not} hold when reasoning about partial subpaths. This prevents the use of prior semi-position free approaches in graphics~\cite{Guo:2018:Positionfree}, which formulate bidirectional estimators only for symmetric extinction coefficients.

However, in our position-free formulation, we eliminate all distance dimensions. This removes the sources of non-reciprocal PDFs related to the asymmetric extinction coefficient, making it possible to implement fully bidirectional estimators without much complication. We give additional details and pseudo-code of our full bidirectional estimator in the supplemental material.

\begin{figure*}[t]
	\centering
	\setlength{\tabcolsep}{0pt}
	\hspace{-5mm}
	\begin{tabular}{c@{\hspace{2mm}}cc@{\hspace{5mm}}cc}
		&\multicolumn{2}{c}{$\theta_i=45^\circ$} & \multicolumn{2}{c}{$\theta_i=80^\circ$}\\
		\cmidrule(lr{5mm}){2-3}\cmidrule(lr){4-5}\\[-2mm]
		\rotatebox[origin=c]{90}{\hspace{5mm}$\alpha=0.2$}&\parbox{0.24\textwidth}{\input{gfx/microfacet-plots-new/MicrofacetReflectance-0.2-45.pgf}}&\parbox{0.24\textwidth}{\input{gfx/microfacet-plots-new/MicrofacetTTUV-0.2-45.pgf}} & \parbox{0.24\textwidth}{\input{gfx/microfacet-plots-new/MicrofacetReflectance-0.2-80.pgf}} & \parbox{0.24\textwidth}{\input{gfx/microfacet-plots-new/MicrofacetTTUV-0.2-80.pgf}} \\
		\rotatebox[origin=c]{90}{\hspace{5mm}$\alpha=0.4$}&\parbox{0.24\textwidth}{\input{gfx/microfacet-plots-new/MicrofacetReflectance-0.4-45.pgf}} & \parbox{0.24\textwidth}{\input{gfx/microfacet-plots-new/MicrofacetTTUV-0.4-45.pgf}} & \parbox{0.24\textwidth}{\input{gfx/microfacet-plots-new/MicrofacetReflectance-0.4-80.pgf}} & \parbox{0.24\textwidth}{\input{gfx/microfacet-plots-new/MicrofacetTTUV-0.4-80.pgf}} \\
		\rotatebox[origin=c]{90}{\hspace{5mm}$\alpha=0.75$}&\parbox{0.24\textwidth}{\input{gfx/microfacet-plots-new/MicrofacetReflectance-0.75-45.pgf}} & \parbox{0.24\textwidth}{\input{gfx/microfacet-plots-new/MicrofacetTTUV-0.75-45.pgf}} & \parbox{0.24\textwidth}{\input{gfx/microfacet-plots-new/MicrofacetReflectance-0.75-80.pgf}} & \parbox{0.24\textwidth}{\input{gfx/microfacet-plots-new/MicrofacetTTUV-0.75-80.pgf}} \\
		\rotatebox[origin=c]{90}{\hspace{5mm}$\alpha=1.0$}&\parbox{0.24\textwidth}{\input{gfx/microfacet-plots-new/MicrofacetReflectance-1.0-45.pgf}} & \parbox{0.24\textwidth}{\input{gfx/microfacet-plots-new/MicrofacetTTUV-1.0-45.pgf}} & \parbox{0.24\textwidth}{\input{gfx/microfacet-plots-new/MicrofacetReflectance-1.0-80.pgf}} & \parbox{0.24\textwidth}{\input{gfx/microfacet-plots-new/MicrofacetTTUV-1.0-80.pgf}} \\
	\end{tabular}
	\includegraphics[width=0.6\textwidth]{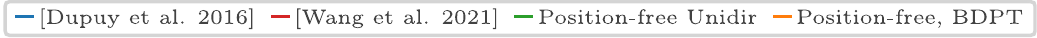}
    \caption{We show mean BSDF value (left half of columns) and inverse efficiency (right half of columns, lower is better) for reflectance from a conductive GGX microfacet with multiple scattering for different roughnesses and incident angles. We compare four methods: The baseline algorithm of Dupuy et al. \cite{Dupuy:2016:Additional} (blue); the approximate position-free algorithm of Wang et al.~\cite{Wang:2021:Position} (red); and our two unbiased position-free algorithms (green and orange). All position-free algorithms significantly outperform the baseline; the method of Wang et al.\ does so at the cost of significant bias. Our methods are unbiased and significantly outperform both the baseline and, at non-normal incidence, also the algorithm of Wang et al. \label{fig:microfacet-efficiency}}
\end{figure*}

\subsubsection{Numerical Stability} We found that, unlike the slab case in \cref{sec:slab-application}, our position-free algorithms for microfacets are numerically stable. This is explained by the nature of multiple scattering on microfacets: The overwhelming majority of energy is contained in low-order bounces (less than 4 segments), and practically no path grows beyond 10 segments (roughly $0.01\%$ at roughness $\alpha=1$; much less at lower roughnesses). 
Therefore, we did not find the need to implement any numerical mitigation strategies such as was required for the slab.

\subsubsection{Russian Roulette} For microfacets, we found it beneficial to perform early termination of low-energy paths. We can easily compute the energy left within the slab from $h_l$, and use this as the termination probability after each collision. We give details in the supplemental.

\subsection{Comparison to Wang et al.}\label{sec:wang}

Wang et al.~\cite{Wang:2021:Position} also propose a fully position-free MC transport model for Smith microfacets using a different methodology. Instead of computing the true $\ExitP$, they suggest instead using products of the monostatic Smith shadowing function $G_1(\Dir)=1/(1+\ExtCoeff_\mathrm{smith}(\Dir))$. This is equivalent to computing $\ExitP$ of a modified random walk, where after each collision the location of the photon is ``scrambled'' and drawn uniformly from the distribution of microsurface heights. In this case, $h_l$ resets to a known distribution ($h_l(z)=\exp^{-z}$) after each collision, and computing $\ExitP$ becomes trivial.

This segment-wise decorrelation of the random walk leads to a biased result. As we will show in the next section, this bias can be significant and is not outweighed by commensurate efficiency gains. However, this approach is a useful fallback strategy for when $\ExitP$ cannot be evaluated analytically, i.e. for refracted paths (\cref{sec:dielectric-microfacets}).

\subsection{Results}

In \cref{fig:microfacet-efficiency}, we show mean and efficiency graphs for evaluating multiple scattering on GGX conductors over a variety of roughnesses and incident inclinations. We compare four different methods. The first is the original algorithm proposed by Heitz et al. \cite{Heitz:2016:Multiple}, which performs MIS of unidirectional path- and light tracing in the semi-infinite slab. This baseline method can be made position-free in two different ways: Either approximately, by replacing distance-related terms with the analytic Smith monostatic shadowing function (this is the method of Wang et al.~\cite{Wang:2021:Position}); or by replacing them with the true closed-form solution, which is our \textsc{Position-free Unidir} strategy. Additionally, we compare against unbiased, fully bidirectional position-free estimation, which is our \textsc{Position-free BDPT} strategy.

Even when taking estimator cost into account, we find that all position-free algorithms consistently beat the baseline estimator of Heitz et al. by significant margin over a wide range of parameters. However, the approximate algorithm of Wang et al. shows significant bias, especially at higher roughnesses. When comparing efficiency, our bidirectional and MIS-of-unidirectional estimators are roughly equivalent: The fully bidirectional estimator exhibits lower variance, but is approximately $2\times$ as expensive.  In a complete renderer, the relative cost of evaluating the bidirectional estimator becomes smaller compared to the cost of tracing a full path, and this estimator becomes much more competitive.
The method of Wang et al. is slightly more efficient at normal incidence, but significantly less so at lower inclinations.

To demonstrate practical rendering results, we compare renderings of rough microfacet conductors of varying roughnesses under direct lighting with our bidirectional, position-free algorithm to that of prior work in \cref{fig:teaser}. Despite nearly identical rendering cost, our algorithm provides significant overall variance reduction. It also eliminates high-intensity fireflies of prior work witnessed e.g.\ on the back wall.

\subsection{Dielectric Microfacets} \label{sec:dielectric-microfacets}

\begin{figure*}[t]
	\centering
	\setlength{\tabcolsep}{0pt}
	\hspace{-5mm}
	\begin{tabular}{c@{\hspace{2mm}}cc@{\hspace{5mm}}cc}
		&\multicolumn{2}{c}{$\theta_i=45^\circ$} & \multicolumn{2}{c}{$\theta_i=80^\circ$}\\
		\cmidrule(lr{5mm}){2-3}\cmidrule(lr){4-5}\\[-2mm]
		\rotatebox[origin=c]{90}{\hspace{5mm}$\alpha=0.2$}&\parbox{0.24\textwidth}{\input{gfx/dielectric-plots-new/MicrofacetReflectance-0.2-45.pgf}}&\parbox{0.24\textwidth}{\input{gfx/dielectric-plots-new/MicrofacetTTUV-0.2-45.pgf}} & \parbox{0.24\textwidth}{\input{gfx/dielectric-plots-new/MicrofacetReflectance-0.2-80.pgf}} & \parbox{0.24\textwidth}{\input{gfx/dielectric-plots-new/MicrofacetTTUV-0.2-80.pgf}} \\
		\rotatebox[origin=c]{90}{\hspace{5mm}$\alpha=0.4$}&\parbox{0.24\textwidth}{\input{gfx/dielectric-plots-new/MicrofacetReflectance-0.4-45.pgf}} & \parbox{0.24\textwidth}{\input{gfx/dielectric-plots-new/MicrofacetTTUV-0.4-45.pgf}} & \parbox{0.24\textwidth}{\input{gfx/dielectric-plots-new/MicrofacetReflectance-0.4-80.pgf}} & \parbox{0.24\textwidth}{\input{gfx/dielectric-plots-new/MicrofacetTTUV-0.4-80.pgf}} \\
		\rotatebox[origin=c]{90}{\hspace{5mm}$\alpha=0.75$}&\parbox{0.24\textwidth}{\input{gfx/dielectric-plots-new/MicrofacetReflectance-0.75-45.pgf}} & \parbox{0.24\textwidth}{\input{gfx/dielectric-plots-new/MicrofacetTTUV-0.75-45.pgf}} & \parbox{0.24\textwidth}{\input{gfx/dielectric-plots-new/MicrofacetReflectance-0.75-80.pgf}} & \parbox{0.24\textwidth}{\input{gfx/dielectric-plots-new/MicrofacetTTUV-0.75-80.pgf}} \\
		\rotatebox[origin=c]{90}{\hspace{5mm}$\alpha=1.0$}&\parbox{0.24\textwidth}{\input{gfx/dielectric-plots-new/MicrofacetReflectance-1.0-45.pgf}} & \parbox{0.24\textwidth}{\input{gfx/dielectric-plots-new/MicrofacetTTUV-1.0-45.pgf}} & \parbox{0.24\textwidth}{\input{gfx/dielectric-plots-new/MicrofacetReflectance-1.0-80.pgf}} & \parbox{0.24\textwidth}{\input{gfx/dielectric-plots-new/MicrofacetTTUV-1.0-80.pgf}} \\
	\end{tabular}
	\includegraphics[width=0.4\textwidth]{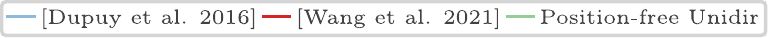}
    \caption{We show mean BSDF value (left half of columns) and inverse efficiency (right half of columns, lower is better) for reflectance from a dielectric GGX microfacet with multiple scattering for different roughnesses and incident angles, and an index of refraction of $1.8$. We compare three methods: The baseline algorithm of Dupuy et al. \cite{Dupuy:2016:Additional} (blue); the approximate position-free algorithm of Wang et al.~\cite{Wang:2021:Position} (red); and our position-free algorithm (green). The efficiency of our method and that of Wang et al. is identical, and both outperform the baseline. However, our method is unbiased for the reflection lobe ($\theta_o \in [-90, 90]$) and restricts bias to the refraction lobe, where it matches that of Wang et al. \label{fig:dielectric-efficiency}}
\end{figure*}

Unfortunately, our theory does not allow for fully unbiased position-free integration of dielectric microfacets.
Even though dielectric multiple scattering can also be expressed via a random walk, the walk can transition between both sides of the micro-surface upon refraction.
In the formulation of Dupuy et al.~\cite{Dupuy:2016:Additional}, this corresponds to a \emph{jump} of the photon, where its depth $z$ is remapped to $z_\mathrm{refract} = \log(1 - e^{-z})$.
To make this compatible with our approach, we would need to find a suitable representation of $h_l(z_\mathrm{refract})$ that can be expressed (and integrated) conveniently.
We have not been able to find such a suitable representation; integrals of this nature result in beta functions, which have resisted much further manipulation.

There are two ways to deal with this problem, which affect either the variance or the bias of the refracted lobe. We could either fall back to the baseline algorithm of Dupuy et al.~\cite{Dupuy:2016:Additional} whenever the random walk refracts; this would lead to unbiased results, but the efficiency in the refracted lobe would be equivalent to the non-position-free baseline. Alternatively, we could fall back to the approach of Wang et al.~\cite{Wang:2021:Position} when the random walk refracts, and maintain the low variance of the position-free approach, at the cost of bias in the refracted lobe. We evaluate the efficiency of the latter approach in \cref{fig:dielectric-efficiency}. As expected, we match the efficiency of the approach of Wang et al., and maintain unbiasedness in the reflected lobe ($\theta_o\in [-90, 90]$). However, we incur the same amount of bias in the refracted lobe.

\section{Conclusion, Limitations and Future Work}\label{sec:conclusion}

We present a new position-free path integral for Monte Carlo estimation of reflectance from homogeneous slabs. We separate out the probability of exiting the slab from other terms in the path integral, and reduce the problem of computing it to manipulating the distribution of heights of a photon in a simplified 1D medium. We show that this distribution reduces to a simple sum of exponentials, the coefficients of which can be computed with simple iterative rules.

We apply our method to two applications in computing slab reflectance and estimating multiple scattering on microfacet conductors. Both are fundamental graphics problems and are directly amenable to our position-free estimation. Our proposed algorithms are simple, and we demonstrate significant variance reduction compared to the baseline in both scenarios.

Our work has a number of limitations and avenues for future research.

\paragraph*{Dielectric Microfacets} Our current formulation cannot achieve fully unbiased position-free integration on dielectric microfacets. This is due to the warping of the height distribution that occurs when the random walk refracts. We propose a scheme that maintains unbiasedness in the reflected lobe, but incurs the same bias as the approach of Wang et al.~\cite{Wang:2021:Position} upon refraction. If it cannot be removed, it would be interesting to explore how bias in the refracted lobe could be reduced. One potential approach could be to optimize the parameters of a new height distribution that most closely matches the warped distribution $h_l(z_\mathrm{refract})$. This would allow for position-free integration even after refraction with reduced bias compared to ignoring the height distribution completely.

\paragraph*{Layered slabs} Another interesting extension of our work would be to natively handle layered, index-matched slabs, such as those of Wang et al.\cite{Wang:2021:Spongecake}. Although we can readily support such use cases by layering individual instances of \cref{sec:slab-application}, further variance reduction may be possible by treating the slabs as a combined medium with strata of different extinction, and tracking a single height distribution over all slabs simultaneously.

\printbibliography

@article{Heitz:2016:Multiple,
    author = {Heitz, Eric and Hanika, Johannes and d'Eon, Eugene and Dachsbacher, Carsten},
    title = {Multiple-Scattering Microfacet BSDFs with the Smith Model},
    year = {2016},
    issue_date = {July 2016},
    publisher = {Association for Computing Machinery},
    address = {New York, NY, USA},
    volume = {35},
    number = {4},
    issn = {0730-0301},
    doi = {10.1145/2897824.2925943},
    journal = {ACM Trans. Graph.},
    month = {7},
    articleno = {58},
    numpages = {14},
}

@article{Papas:2014:Paper,
    author = {Papas, Marios and de Mesa, Krystle and Jensen, Henrik Wann},
    title = {A Physically-Based BSDF for Modeling the Appearance of Paper},
    journal = {Computer Graphics Forum},
    volume = {33},
    number = {4},
    pages = {133-142},
    doi = {10.1111/cgf.12420},
    year = {2014}
}

@article{Wang:2021:Spongecake,
  title = {SpongeCake: A Layered Microflake Surface Appearance Model},
  author = {Beibei Wang and Wenhua Jin and Milo\u{s} Ha\u{s}an and Ling-Qi Yan},
  year = {2021},
  month = oct,
  archivePrefix = {arXiv},
  eprint = {2110.07145},
  eprinttype = {arxiv},
  journal = {arXiv:2110.07145 [cs.GR]},
  primaryClass = {cs.GR}
}

@article{Sears:1975:Slow,
  title={Slow-neutron multiple scattering},
  author={Sears, Varley F},
  journal={Advances in Physics},
  volume={24},
  number={1},
  pages={1--45},
  year={1975},
  publisher={Taylor \& Francis},
  url={https://doi.org/10.1080/00018737500101361}
}

@inproceedings{ stam01,
   author = { Jos Stam },
   booktitle = { Rendering Techniques },
   title = { An Illumination Model for a Skin Layer Bounded by Rough Surfaces },
   pages = { 39--52 },
   year = { 2001 },
   url={https://doi.org/10.1007/978-3-7091-6242-2_4}
}

@article{Berger:1956:Reflection,
  title={Reflection and Transmission of Gamma Radiation by Barriers: Semianalytic Monte Carlo Calculation},
  author={Berger, Martin J and Doggett, John},
  journal={Journal of Research of the National Bureau of Standards},
  volume={56},
  number={2},
  pages={89},
  year={1956},
  publisher={National Bureau of Standards}
}

@article{Wang:2021:Position,
  author    = {Beibei Wang and
               Wenhua Jin and
               Jiahui Fan and
               Jian Yang and
               Nicolas Holzschuch and
               Ling{-}Qi Yan},
  title     = {Position-free Multiple-bounce Computations for Smith Microfacet BSDFs},
  journal   = {CoRR},
  volume    = {abs/2109.14398},
  year      = {2021},
  url       = {https://arxiv.org/abs/2109.14398},
  eprinttype = {arXiv},
  eprint    = {2109.14398},
  bibsource = {dblp computer science bibliography, https://dblp.org}
}

@article{Chilton:1969:Imaging,
  title={Imaging properties of light scattered by the sea},
  author={Chilton, Frank and Jones, Dixon D and Talley, Wilson K},
  journal={JOSA},
  volume={59},
  number={8},
  pages={891--898},
  year={1969},
  publisher={Optical Society of America},
  url={https://doi.org/10.1364/JOSA.59.000891}
}

@article{Drawbaugh:1961:Solution,
  title={{On the solution of transport problems by conditional Monte Carlo}},
  author={Drawbaugh, DW},
  journal={Nuclear Sci. and Eng.},
  volume={9},
  year={1961},
  publisher={Combustion Engineering, Inc., Windsor, Conn.},
  url={https://doi.org/10.13182/NSE61-A15603}
}

@techreport{Amster:1960:Euripus,
  title={{EURIPUS-3 and DAEDALUS--Monte Carlo Density Codes for the IBM-704}},
  author={Amster, Harvey J and Kuehn, Heidi G and Spanier, Jerome},
  year={1960},
  institution={Westinghouse Electric Corp. Bettis Atomic Power Lab., Pittsburgh},
  number={WAPD-TM-205}
}

@article{Amster:1964:Spatial,
  title={{Spatial-Distribution Functions for Calculating Neutron Densities by Monte Carlo}},
  author={Amster, Harvey J and Talley, Wilson K},
  journal={Nuclear Science and Engineering},
  volume={20},
  number={1},
  pages={53--59},
  year={1964},
  publisher={Taylor \& Francis},
  url={https://doi.org/10.13182/NSE64-A19274}
}

@techreport{Kahn:1949:Stochastic,
  title={{Stochastic (Monte Carlo) attenuation analysis}},
  author={Kahn, Herman},
  year={1949},
  number={P-88},
  institution={RAND CORP, SANTA MONICA CALIFORNIA}
}

@techreport{ goertzel1950proposed,
  title={A Proposed Particle Attenuation Problem},
  author={Goertzel, Gerald},
  number={AECD-2808},
  year={1950},
  institution={Oak Ridge National Laboratory}
}

@article{Plass:1968:Monte,
  title={{Monte Carlo calculations of light scattering from clouds}},
  author={Plass, Gilbert N and Kattawar, George W},
  journal={Applied optics},
  volume={7},
  number={3},
  pages={415--419},
  year={1968},
  publisher={Optical Society of America},
  url={https://doi.org/10.1364/AO.7.000415}
}

@inproceedings{ xia2020gaussian,
  title={Gaussian Product Sampling for Rendering Layered Materials},
  author={Xia, Mengqi and Walter, Bruce and Hery, Christophe and Marschner, Steve},
  booktitle={Computer Graphics Forum},
  volume={39},
  number={1},
  pages={420--435},
  year={2020},
  organization={Wiley Online Library},
  url={https://doi.org/10.1111/cgf.13883}
}

@inproceedings{ gamboa2020efficient,
  title={An Efficient Transport Estimator for Complex Layered Materials},
  author={Gamboa, Luis E and Gruson, Adrien and Nowrouzezahrai, Derek},
  booktitle={Computer Graphics Forum},
  volume={39},
  number={2},
  pages={363--371},
  year={2020},
  organization={Wiley Online Library},
  url={https://doi.org/10.1111/cgf.13936}
}

@inproceedings{ weidlich2007arbitrarily,
  title={Arbitrarily layered micro-facet surfaces},
  author={Weidlich, Andrea and Wilkie, Alexander},
  booktitle={Proceedings of the 5th international conference on Computer graphics and interactive techniques in Australia and Southeast Asia},
  pages={171--178},
  year={2007}
}

@misc{Bitterli:2018:Tungsten,
  title = {Tungsten Renderer},
  author = {Bitterli, Benedikt},
  year = {2018},
  url = {https://github.com/tunabrain/tungsten/}
}

@article{Blinn:1982:Light,
  title = {Light Reflection Functions for Simulation of Clouds and Dusty Surfaces},
  author = {Blinn, James F.},
  year = {1982},
  month = jul,
  volume = {16},
  pages = {21--29},
  issn = {0097-8930},
  doi = {10/dz2jzw},
  journal = CG_SIGGRAPH75-92,
  number = {3}
}

@article{Donner:2005:Light,
  title = {Light Diffusion in Multi-Layered Translucent Materials},
  author = {Donner, Craig and Jensen, Henrik Wann},
  year = {2005},
  month = jul,
  volume = {24},
  pages = {1032--1039},
  issn = {0730-0301},
  doi = {10/chbmgw},
  journal = TOG_SIGGRAPH02,
  number = {3}
}

@inproceedings{Dupuy:2016:Additional,
  title = {Additional Progress towards the Unification of Microfacet and Microflake Theories},
  booktitle = PROC_EGSR_EII,
  author = {Dupuy, Jonathan and Heitz, Eric and {d'Eon}, Eugene},
  year = {2016},
  pages = {55--63},
  publisher = PubEG,
  doi = {10/gfz72z},
  isbn = {978-3-03868-019-2}
}

@article{Dwivedi:1982:New,
  title = {A New Importance Biasing Scheme for Deep-Penetration {{Monte Carlo}}},
  author = {Dwivedi, S. R.},
  year = {1982},
  month = jan,
  volume = {9},
  pages = {359--368},
  issn = {0306-4549},
  doi = {10/czbz26},
  journal = AnNE,
  number = {7}
}

@article{Farrell:1992:Diffusion,
  title = {A Diffusion Theory Model of Spatially Resolved, Steady-State Diffuse Reflectance for the Noninvasive Determination of Tissue Optical Properties in Vivo},
  author = {Farrell, T. J. and Patterson, Michael S. and Wilson, B.},
  year = {1992},
  volume = {19},
  pages = {879--888},
  doi = {10/b9rwpz},
  journal = {Medical Physics},
  number = {4}
}

@article{Guo:2018:Positionfree,
  title = {Position-Free {{Monte Carlo}} Simulation for Arbitrary Layered {{BSDFs}}},
  author = {Guo, Yu and Ha{\v s}an, Milo{\v s} and Zhao, Shuang},
  year = {2018},
  month = dec,
  volume = {37},
  pages = {279:1--279:14},
  issn = {0730-0301},
  doi = {10/db3c},
  journal = TOG_SIGGRAPHAsia,
  number = {6}
}

@inproceedings{Hanrahan:1993:Reflection,
  title = {Reflection from Layered Surfaces Due to Subsurface Scattering},
  booktitle = PROC_SIGGRAPH93-01,
  author = {Hanrahan, Pat and Krueger, Wolfgang},
  year = {1993},
  pages = {165--174},
  publisher = PubACM,
  address = {{New York, NY, USA}},
  doi = {10/b4tw3j},
  isbn = {978-0-89791-601-1}
}

@article{Jakob:2014:Comprehensive,
  title = {A Comprehensive Framework for Rendering Layered Materials},
  author = {Jakob, Wenzel and {d'Eon}, Eugene and Jakob, Otto and Marschner, Steve},
  year = {2014},
  month = jul,
  volume = {33},
  pages = {118:1--118:14},
  issn = {0730-0301},
  doi = {10/f6cpsq},
  journal = TOG_SIGGRAPH02,
  number = {4}
}

@book{Pharr:2016:Physically,
  title = {Physically Based Rendering: {{From}} Theory to Implementation},
  shorttitle = {Physically Based Rendering},
  author = {Pharr, Matt and Jakob, Wenzel and Humphreys, Greg},
  year = {2016},
  edition = {3rd},
  publisher = PubMK,
  address = {{Cambridge, MA}},
  isbn = {978-0-12-800645-0},
  lccn = {T385 .P486 2017}
}

@inproceedings{Veach:1995:Optimally,
  title = {Optimally Combining Sampling Techniques for {{Monte Carlo}} Rendering},
  booktitle = PROC_SIGGRAPH93-01,
  author = {Veach, Eric and Guibas, Leonidas J.},
  year = {1995},
  month = aug,
  volume = {29},
  pages = {419--428},
  publisher = PubACM,
  doi = {10/d7b6n4},
  isbn = {978-0-89791-701-8}
}

@inproceedings{Walter:2007:Microfacet,
  title = {Microfacet Models for Refraction through Rough Surfaces},
  booktitle = RT_EGSR03-07,
  author = {Walter, Bruce and Marschner, Stephen R. and Li, Hongsong and Torrance, Kenneth E.},
  year = {2007},
  month = jun,
  pages = {195--206},
  publisher = PubEG,
  address = {{Grenoble, France}},
  doi = {10/gfz4kg},
  isbn = {978-3-905673-52-4},
  numpages = {12}
}

@string{CG_SIGGRAPH75-92         = "Computer Graphics (Proceedings of {SIGGRAPH})"}

@string{PROC_SIGGRAPH93-01       = "Annual Conference Series (Proceedings of {SIGGRAPH})"}

@string{TOG_SIGGRAPH02           = "{ACM} Transactions on Graphics (Proceedings of {SIGGRAPH})"}

@string{TOG_SIGGRAPHAsia         = "{ACM} Transactions on Graphics (Proceedings of {SIGGRAPH Asia})"}

@string{RT_EGSR03-07   = "Rendering Techniques (Proceedings of the Eurographics Symposium on Rendering)"}

@string{PROC_EGSR_EII  = "Proceedings of EGSR (Experimental Ideas \& Implementations)"}

@string{JOSA        = "Journal of the Optical Society of America"}

@string{AnNE      = "Annals of Nuclear Energy"}

@string{JOSA      = "Journal of the Optical Society of America"}

@STRING{PubACM       = "ACM Press"}

@STRING{PubEG        = "Eurographics Association"}

@STRING{PubMK        = "Morgan Kaufmann"}
\end{document}